\documentclass[12pt]{article}
\usepackage{epsfig}
\usepackage{amsfonts}
\usepackage{latexsym}
\usepackage{amsmath,amssymb}
\usepackage{mathrsfs}
\usepackage{hyperref}
\usepackage{setspace}
\usepackage{color}
\usepackage{bm}
\numberwithin{equation}{section}

\begin{document}

\begin{titlepage}
\unitlength = 1mm

\vskip 1cm
\begin{center}

 {\Large {\textsc{\textbf{A note on initial state entanglement\\ in inflationary cosmology}}}}

\vspace{1.8cm}
Sugumi Kanno

\vspace{1cm}

{\it $^*$ Department of Theoretical Physics and History of Science,
University of the Basque Country UPV/EHU,
48080 Bilbao, Spain}

\vspace{0.2cm}

{\it $^\flat$ IKERBASQUE, Basque Foundation for Science, 
Maria Diaz de Haro 3,
48013, Bilbao, Spain}

\vspace{0.2cm}

\vskip 1.5cm

\begin{abstract}
\baselineskip=6mm
We give a new interpretation of the effect of initial state entanglement on the spectrum of vacuum fluctuations. We consider an initially entangled state between two free massive scalar fields in de Sitter space. We construct the initial state by making use of a Bogoliubov transformation between the Bunch-Davies vacuum and a four mode squeezed state, and then derive the exact power spectrum for one of the scalar fields. We demonstrate that an oscillatory spectrum hardly appears for the initially entangled state unless an ad hoc absolute value of the Bogoliubov coefficients is chosen.
\end{abstract}

\vspace{1.0cm}

\end{center}
\end{titlepage}

\pagestyle{plain}
\setcounter{page}{1}
\newcounter{bean}
\baselineskip18pt

\setcounter{tocdepth}{2}

\tableofcontents

\section{Introduction}
Historically, quantum entanglement has been one of the most fascinating features predicted by quantum mechanics since Einstein-Podolsky-Rosen (EPR) pointed out that performing a local measurement may affect the outcome of local measurements instantaneously beyond the light cone in 1935~\cite{Einstein:1935rr}. After a convincing test that the quantum entanglement is a fundamental aspect of quantum mechanics by measuring correlations of linear polarizations of pairs of photons~\cite{Aspect:1981zz, Aspect:1982fx} was performed by Aspect et al in 1981, more attention has been paid to how to make use of quantum entanglement of EPR pairs in quantum cryptography and quantum teleportation (see \cite{Horodecki:2009zz} and references therein).

Entanglement entropy has now been established as a suitable measure of the degree of entanglement of a quantum system. It has become a useful tool in quantum field theory to characterize the nature of long range correlations and been developed in condensed matter physics, quantum information and high energy physics. Recently, to discuss a gravity dual of quantum entanglement in de Sitter space, Maldacena and Pimentel developed an explicit method to calculate the entanglement entropy in a quantum field theory in the Bunch-Davies vacuum of the de Sitter space~\cite{Maldacena:2012xp}. They showed that quantum entanglement can exist between two causally
disconnected open charts in de Sitter space. Since the interior of a nucleated
bubble can be described by the open chart~\cite{Coleman:1980aw}, their result infers that the regions inside and outside of the bubble are entangled. If we regard the two open charts as two separated bubbles, the two bubbles are entangled. Then, the spectrum of vacuum fluctuations in one of the open charts was examined in \cite{Kanno:2014ifa} and it was found that the quantum entanglement affects the shape of the spectrum on large scales comparable to or greater than the curvature radius, whereas the spectrum on small scales remains intact~\cite{Kanno:2014ifa, Dimitrakopoulos:2015yva}. This is because the Bogoliubov transformation between the Bunch-Davies vacuum and the vacua of the open charts depends on the wavenumber and vanishes in the small scale limit. Thus there is no entanglement on small scales. This result is consistent with the fact that there should
be no difference between the open chart vacuum and the Bunch-Davies vacuum, since spacetime looks flat in the small scale limit. In other words, this result that there is no entanglement on small scales is reasonable as long as bubbles are nucleated
in vacuum, with no particle production on small scales.

The research of~\cite{Albrecht:2014aga}, however, discussed that oscillatory spectra may appear as the effect of quantum entanglement even on small scales if we consider an initially entangled state between two free massive scalar fields in de Sitter space. If the amplitude of oscillations is small enough, the spectrum cannot be distinguished from a flat spectrum. On the other hand, the research of~\cite{Kanno:2014bma} studied two de Sitter spaces instead of the two scalar fields and showed that an initial state entanglement between two causally disconnected de Sitter spaces may remain on small scales. In the bubble universes or so called multiverse, an infinite number of universes can be produced once inflation happens~\cite{Sato:1981gv, Vilenkin:1983xq, Linde:1986fc, Linde:1986fd, Bousso:2000xa, Susskind:2003kw}. In such a situation, some bubbles may be nucleated as entangled pairs of bubbles and their initial states may be entangled with their partner bubbles\footnote{The nucleation of bubbles is analogous to Schwinger pair production~\cite{Schwinger:1951nm, Brown:1988kg, Garriga:2012qp, Garriga:2013pga, Frob:2014zka}.}. Then, some effects of the entanglement with unobservable bubbles may appear in the CMB of our bubble. In fact, the research of~\cite{Kanno:2015lja} showed that the quantum entanglement may be imprinted on the cosmological observables.

Since the initially entangled state considered in~\cite{Kanno:2015lja} corresponds to truncation of a squeezed state of pairs of $n$-particles, the degree of entanglement is represented by the Bogoliubov coefficients between the Bunch-Davies vacuum and the squeezed state. Thus one may expect to obtain oscillatory spectra with an appropriate choice of the Bogoliubov coefficients $\beta$ as demonstrated in~\cite{Albrecht:2014aga}, although such oscillatory spectra are not solely due to the entanglement because initially non-entangled states also produce the oscillatory spectra because of the Bogoliubov coefficients $\beta$~\cite{Kanno:2015lja}. The research of~\cite{Kanno:2015lja} showed that the relative phase between the coefficients $\alpha$ and $\beta$ due to quantum interference may distinguish the spectrum of initially entangled states from the one of non-entangled.

The computation in~\cite{Albrecht:2014aga} is done in the Schr\"odinger formalism, so it is not easy to see the physical origin of the oscillatory behavior in the spectrum. In this paper, we instead adopt the Heisenberg formalism and clarify what causes the oscillatory behavior in the spectrum when the initial state is entangled. We then give a new interpretation of the result.

\section{Four mode squeezed state in de Sitter space}
\label{s2}
In~\cite{Albrecht:2014aga}, an initially four mode squeezed state of two free massive scalar fields in de Sitter space was discussed in the Schr\"{o}dinger formalism. As a result, oscillatory spectra appeared. As we show below, this state describes an entangled state between the two scalar fields. Here, we discuss their system in the Heisenberg formalism and try to figure out what caused the oscillatory behavior in the spectrum.

\subsection{The set-up}
We consider two free massive scalar fields $\phi_{\bm k}$ and $\chi_{\bm k}$ in de Sitter with the action~\cite{Albrecht:2014aga}
\begin{eqnarray}
S=\frac{1}{2}\int d\eta \sum_{\bm k}\left[\,
a^2\left(\phi^\prime_{\bm k}\phi^{*\prime}_{\bm k}
-k^2\phi_{\bm k}\phi_{\bm k}^{*}\right)
-a^4m_{\phi}^2\,\phi_{\bm k}\phi_{\bm k}^{*}
+a^2\left(\chi^\prime_{\bm k}\chi^{*\prime}_{\bm k}
-k^2\chi_{\bm k}\chi_{\bm k}^{*}\right)
-a^4m_{\chi}^2\,\chi_{\bm k}\chi_{\bm k}^{*}
\,\right]\,,
\end{eqnarray}
where the scale factor $a=-1/(H\eta)$ and a prime denotes derivative with respect to the conformal time $\eta$. ${\bm k}$ is the wave vector.
The canonical conjugate momentum for $\phi_{\bm k}$ and $\chi_{\bm k}$ is give by
\begin{eqnarray}
P_{\phi\,{\bm k}}=\frac{\partial S}{\partial\phi_{\bm k}^\prime}=a^2\phi_{\bm k}^{*\prime}\,,\qquad
P_{\chi\,{\bm k}}=\frac{\partial S}{\partial\chi_{\bm k}^\prime}=a^2\chi_{\bm k}^{*\prime}\,.
\label{cm1}
\end{eqnarray}
Each field is expanded as
\begin{eqnarray}
\phi_{\bm k}&=&a_{\bm k}u_k(\eta)+a_{-{\bm k}}^\dag u_k^*(\eta)\,,
\qquad \left[a_{\bm k},a_{\bm p}^\dag\right]
=\delta_{\bm k, \bm p}\label{phiau}\,,\nonumber\\
\chi_{\bm k}&=&b_{\bm k}v_k(\eta)+b_{-{\bm k}}^\dag v_k^*(\eta)\,,\qquad\,\, \left[b_{\bm k},b_{\bm p}^\dag\right]=\delta_{\bm k,\bm p}\,,
\label{phichi1}
\end{eqnarray}
where $k$ is the magnitude of the wave vector ${\bm k}$. The different masses of $m_\phi$ and $m_\chi$ produce different mode functions $u_k$ and $v_k$, which satisfy
\begin{eqnarray}
&&u_k^{\prime\prime}+\left(k^2+m_\phi^2\,a^2-\frac{a^{\prime\prime}}{a}\right)u_k=0\,,
\qquad
u_ku_k^{*\prime}-u_k^{*}u_k^\prime=\frac{i}{a^2}\,,\nonumber\\
&&v_k^{\prime\prime}+\left(k^2+m_\chi^2\,a^2-\frac{a^{\prime\prime}}{a}\right)v_k=0\,,
\qquad
v_kv_k^{*\prime}-v_k^{*}v_k^\prime=\frac{i}{a^2}\,.
\label{uv}
\end{eqnarray}
where we used the canonical conjugate relations $[\phi_{\bm k},P_{\phi\,{\bm p}}]=i\delta_{\bm k,\bm p}$, $[\chi_{\bm k},P_{\chi\,{\bm p}}]=i\delta_{\bm k,\bm p}$ for the Klein-Gordon normalization respectively.
The mode functions that realize the Bunch-Davies vacuum at $\eta\rightarrow -\infty$ are respectively given by
\begin{eqnarray}
u_k&=&\frac{\sqrt{\pi}}{2}\sqrt{-\eta}H_{\nu_\phi}^{(1)}(-k\eta)\,,\qquad
\nu_\phi=\sqrt{\frac{9}{4}-\frac{m_\phi^2}{H^2}}\,,\nonumber\\
v_k&=&\frac{\sqrt{\pi}}{2}\sqrt{-\eta}H_{\nu_\chi}^{(1)}(-k\eta)\,,\qquad
\nu_\chi=\sqrt{\frac{9}{4}-\frac{m_\chi^2}{H^2}}\,.
\end{eqnarray}
Note that we assumed $m_\phi^2, m_\chi^2 < 9/4 H^2$ for simplicity. Then the canonical conjugate momentum Eq.~(\ref{cm1}) is expressed as
\begin{eqnarray}
P_{\phi\,{\bm k}}=a^2\left(a_{-{\bm k}}u_k^\prime
+a_{\bm k}^\dag u_k^{*\prime}\right)\,,\qquad
P_{\chi\,{\bm k}}=a^2\left(b_{-{\bm k}}v_k^\prime
+b_{\bm k}^\dag v_k^{*\prime}\right)\,.
\label{cm2}
\end{eqnarray}
By using Eqs.~(\ref{phichi1}) and (\ref{cm2}), the annihilation and creation operators are written as
\begin{eqnarray}
&&a_{\bm k}=u_k^*\frac{\partial}{\partial\phi_{-{\bm k}}}
-ia^2u_k^{*\prime}\phi_{\bm k}\,,\qquad
a_{-{\bm k}}^\dag=-u_k\frac{\partial}{\partial\phi_{-{\bm k}}}
+ia^2u_k^{\prime}\phi_{\bm k}\,,\nonumber\\
&&b_{\bm k}=v_k^*\frac{\partial}{\partial\chi_{-{\bm k}}}
-ia^2v_k^{*\prime}\chi_{\bm k}\,,\qquad\,\,
b_{-{\bm k}}^\dag=-v_k\frac{\partial}{\partial\chi_{-{\bm k}}}
+ia^2v_k^{\prime}\chi_{\bm k}\,.
\label{ab}
\end{eqnarray}
The vacuum state is the state annihilated by both $a_k$ and $b_k$. So if we denote the vacuum for $\phi_{\bm k}$ by $|0\rangle_\phi$ and for $\chi_{\bm k}$ by $|0\rangle_\chi$, the vacuum for the total system is
\begin{eqnarray}
|0\rangle=|0\rangle_\phi|0\rangle_\chi\,.
\end{eqnarray}

\subsection{The Bogoliubov transformation}
Let us consider a state $|\psi\rangle$ defined by Bogoliubov transformations that mix the operators $a_{\bm k}$ with $b_{\bm k}$
\begin{eqnarray}
\tilde{a}_{\bm k}=\alpha_ka_{\bm k}+\beta_kb_{-{\bm k}}^\dag\,,\qquad
\tilde{b}_{\bm k}=\alpha_kb_{\bm k}+\beta_ka_{-{\bm k}}^\dag\,,
\label{tildeab1}
\end{eqnarray}
that is
\begin{eqnarray}
\tilde{a}_{\bm k}|\psi\rangle=\tilde{b}_{\bm k}|\psi\rangle=0\,,
\label{vacuum}
\end{eqnarray}
where 
\begin{eqnarray}
|\alpha_k|^2-|\beta_k|^2=1.
\label{coefficients}
\end{eqnarray}
This state $|\psi\rangle$ is then written by
\begin{eqnarray}
|\psi\rangle=N\exp\left[-\sum_{\bm k}\frac{\beta_k}{\alpha_k}a_{\bm k}^\dag b_{-\bm k}^\dag\right]|0\rangle\,,
\label{heisenberg}
\end{eqnarray}
where $N$ is the normalization factor. This describes a squeezed state of pairs of $n$-particles between the two scalar fields $\phi_{\bm k}$ and $\chi_{\bm k}$
If we expand the exponent in $|\psi\rangle$, we have
\begin{eqnarray}
|\psi\rangle&=&N\left(1-\frac{\beta_k}{\alpha_k}a_{\bm k}^\dag b_{-\bm k}^\dag
+\cdots\right)|0\rangle\nonumber\\
&=&N\left(\,|0\rangle_\phi|0\rangle_\chi-\frac{\beta_k}{\alpha_k}|1\rangle_\phi|1\rangle_\chi+\frac{1}{2}\left(\frac{\beta_k}{\alpha_k}\right)^2|2\rangle_\phi|2\rangle_\chi+\cdots\right)\,.
\end{eqnarray}
This is an entangled state of the ${\cal H}_{\phi}\otimes{\cal H}_{\chi}$ space. If we truncate the above at the one particle state order, we get a standard description of composite system of an entangled state between the vacuum and one-particle states. Thus the degree of entanglement corresponds to the Bogoliubov coefficients and it may depend on the wavenumber $k$. 

\subsection{The spectrum in the Heisenberg picture}

Now, we calculate the spectrum of one of the scalar fields $\phi_{\bm k}$ in the state 
$|\psi\rangle$ in Eq.~(\ref{heisenberg}). Eq.~(\ref{phiau}) and its complex conjugate are expanded in terms of $\tilde{a}_{\bm k}$ and $\tilde{b}_{\bm k}$ as
\begin{eqnarray}
\phi_{\bm k}&=&\left(\alpha_k^*\tilde{a}_{\bm k}-\beta_k\tilde{b}_{-{\bm k}}^\dag\right)u_k+\left(\alpha_k\tilde{a}_{-{\bm k}}^\dag-\beta_k^*\tilde{b}_{\bm k}\right)u_k^*\,,\\
\phi_{\bm k}^\dag&=&\left(\alpha_k\tilde{a}_{\bm k}^\dag-\beta_k^*\tilde{b}_{-{\bm k}}\right)u_k^*+\left(\alpha_k^*\tilde{a}_{-{\bm k}}-\beta_k^*\tilde{b}_{\bm k}^\dag\right)u_k\,.
\end{eqnarray}
Then the spectrum of vacuum fluctuations is calculated as
\begin{eqnarray}
\langle\psi|\phi_{\bm k}\phi_{\bm k}^\dag|\psi\rangle=
|u_k|^2\left(|\alpha_k|^2+|\beta_k|^2\right)
=|u_k|^2\left(1+2|\beta_k|^2\right)\,,
\label{exact1}
\end{eqnarray}
where we used Eq.~(\ref{vacuum}). This is the exact spectrum for the field $\phi_{\bm k}$. Interestingly, we find that the mode function $v_k$ does not come in the spectrum as far as we focus on the spectrum for $\phi_{\bm k}$.
Thus, the possible origin of oscillatory spectra has to be in the absolute value of the Bogoliubov coefficients $|\beta_k|$.

\subsection{The spectrum in the Schr\"odinger picture}
Let us assume that $|\psi\rangle$ in the Schr\"odinger picture is expressed of the form ~\cite{Albrecht:2014aga}
\begin{eqnarray}
\psi=N\exp\left[-\frac{1}{2}\sum_{\bm k}A_k\,\phi_{\bm k}\phi_{-{\bm k}}
-\frac{1}{2}\sum_{\bm k}B_k\,\chi_{\bm k}\chi_{-{\bm k}}
-\frac{1}{2}\sum_{\bm k}C_k\left(\phi_{\bm k}\chi_{-{\bm k}}
+\chi_{\bm k}\phi_{-{\bm k}}\right)\right]\,.
\label{wavefunction}
\end{eqnarray}
The last term proportional to $C_k$ corresponds to an initial state entanglement between $\phi_{\bm k}$ and $\chi_{\bm k}$. Here we note that $\phi_{-\bm k}=\phi_{\bm k}^\dag$ in the Heisenberg picture. So in the Schr\"{o}dinger picture or in the coordinate representation, $\phi_{-\bm k}=\phi_{\bm k}^*$.

Now we try to check if the state $\psi$ in the Schr\"odinger picture (\ref{wavefunction}) corresponds to $|\psi\rangle$ in the Heisenberg picture (\ref{heisenberg}).
Plugging Eq.~(\ref{ab}) into Eq.~(\ref{tildeab1}), we find
\begin{eqnarray}
&&\tilde{a}_{\bm k}=
\alpha_k\left(u_k^*\frac{\partial}{\partial\phi_{-{\bm k}}}
-ia^2u_k^{*\prime}\phi_{\bm k}\right)
+\beta_k\left(-v_k\frac{\partial}{\partial\chi_{-{\bm k}}}
+ia^2v_k^{\prime}\chi_{\bm k}\right)\,,\nonumber\\
&&\tilde{b}_{\bm k}=
\alpha_k\left(v_k^*\frac{\partial}{\partial\chi_{-{\bm k}}}
-ia^2v_k^{*\prime}\chi_{\bm k}\right)
+\beta_k\left(-u_k\frac{\partial}{\partial\phi_{-{\bm k}}}
+ia^2u_k^{\prime}\phi_{\bm k}\right)
\,.
\label{tildeab2}
\end{eqnarray}
If we operate Eq.~(\ref{tildeab2}) on Eq.~(\ref{wavefunction}) and use Eq.~(\ref{vacuum}), we get a system of four homogeneous equations
\begin{eqnarray}
&&A_ku_k^*-\frac{\beta_k}{\alpha_k}C_kv_k=-ia^2u_k^{*\prime}\,,\qquad
B_kv_k-\frac{\alpha_k}{\beta_k}C_ku_k^*=-ia^2v_k^\prime\,,\\
&&B_kv_k^*-\frac{\beta_k}{\alpha_k}C_ku_k=-ia^2v_k^{*\prime}\,,\qquad
A_ku_k-\frac{\alpha_k}{\beta_k}C_kv_k^*=-ia^2u_k^\prime\,.
\end{eqnarray}
Note that one of them is redundant. Those equations determine coefficients $A_p$, $B_p$ and $C_p$ in Eq.~(\ref{wavefunction}) as
\begin{eqnarray}
A_k&=&\frac{1}{|u_k|^2}\frac{\beta_k^2u_kv_k}{\alpha_k^2u_k^*v_k^*-\beta_k^2u_kv_k}
-ia^2\frac{u_k^{*\prime}}{u_k^*}\,,
\label{Ap}\\
B_k&=&\frac{1}{|v_k|^2}\frac{\beta_k^2u_kv_k}{\alpha_k^2u_k^*v_k^*-\beta_k^2u_kv_k}
-ia^2\frac{v_k^{*\prime}}{v_k^*}\,,
\label{Bp}\\
C_k&=&\frac{\alpha_k\beta_k}{\alpha_k^2u_k^*v_k^*-\beta_k^2u_kv_k}\,,
\label{Cp}
\end{eqnarray}
where the relation of the Klein-Gordon normalization in (\ref{uv}) is used. Thus we find that both expressions in the Heisenberg and Schr\"odinger pictures Eqs.~(\ref{heisenberg}) and (\ref{wavefunction}) completely agree with each other when the above (\ref{Ap}), (\ref{Bp}) and (\ref{Cp}) are satisfied.

As we see Eqs.~(\ref{Ap})$\sim$(\ref{Cp}), those expressions look very complicated. In~\cite{Albrecht:2014aga} , they tried to derive Eqs.~(\ref{Ap})$\sim$(\ref{Cp}) approximately, and then calculated the power spectrum in the Schr\"odinger picture. So we try to check if the power spectrum agrees with Eq.~(\ref{exact1}). In the Schr\"odinger picture, the spectrum becomes~\cite{Albrecht:2014aga}
\begin{eqnarray}
\langle\phi_{\bm k}\,\phi_{-{\bm k}}\rangle
=\frac{\left({\rm Re}B_k\right)}{2\{\left({\rm Re}A_k\right)\left({\rm Re}B_k\right)-\left({\rm Re}\,C_k\right)^2\}}\,,
\label{exact2}
\end{eqnarray}
where ${\rm Re}A_k$ is the real part of $A_k$ and similarly for $B_k$ and $C_k$. 

Now we express each variable in $A_k$, $B_k$ and $C_k$ in the form\footnote{Without loss of generality, we assume $\alpha_k$ is real.}
\begin{eqnarray}
\beta_k=b_k+id_k\,,\quad
u_k=x_k+iy_k\,,\quad
v_k=z_k+iw_k\,,
\end{eqnarray}
where $b_k, d_k, x_k, y_k, z_k$ and $w_k$ are real. We plug the above into Eqs.~(\ref{Ap}), (\ref{Bp}) and (\ref{Cp}), and calculate Eq.~(\ref{exact2}). If we expand it in $\beta_k$ up to the second order, we find that the Eq.~(\ref{exact2}) is written by
\begin{eqnarray}
\frac{1}{2a^2}\frac{x^2+y^2}{x^\prime y-y^\prime x}
\left[1-\frac{1}{a^2}\frac{1}{x^\prime y-y^\prime x}\frac{AX+BY}{X^2+Y^2}
+\frac{\alpha^2}{a^4}
\frac{(x^2+y^2)(z^2+w^2)}{(x^\prime y-y^\prime x)(z^\prime w-w^\prime z)}
\left(\frac{bX+dY}{X^2+Y^2}\right)^2\right]\,,
\end{eqnarray}
where we omitted the index $k$ for simplicity and defined
\begin{eqnarray}
&&X=\alpha^2\left(xz-yw\right)\,,\hspace{1.1cm}
A=\left(b^2-d^2\right)\left(xz-yw\right)-2bd\left(yz+xw\right)\,,\nonumber\\
&&Y=-\alpha^2\left(yz+xw\right)\,,\qquad
B=2bd\left(xz-yw\right)+\left(b^2-d^2\right)\left(yz+xw\right)\,.
\end{eqnarray}
If we use the Wronskian $yx'-xy'=1/(2a^2)$, and rewrite the above in terms of $\alpha_k, \beta_k, u_k, v_k$ such as $X^2+Y^2=\alpha^4|u_k|^2|v_k|^2$, the spectrum~(\ref{exact2}) is then expressed as
\begin{eqnarray}
\langle\phi_{\bm k}\,\phi_{-{\bm k}}\rangle
=|u_k|^2\left(1+2|\beta_k|^2+{\cal O}(\beta_k^4)\right)\,.
\end{eqnarray}
We see that the expansion up to the second order of $\beta_k$ completely agrees with the exact spectrum in Eq.~(\ref{exact1}). 
Thus, we find that the degree of freedom of an oscillation in the spectrum appeared in Eq.~(\ref{exact2}) is encoded in the absolute value of the Bogoliubov coefficients $|\beta_k|$ and then the oscillation can appear with an appropriate choice of the $|\beta_k|$. In fact, in~\cite{Albrecht:2014aga}, they choose the phase of $\rho_k$ to be zero and which corresponds to a particular (rather ad hoc) choice of $|\beta_k|$.

\section{Summary and discussion}
\label{s6}

We studied the spectrum of vacuum fluctuations of an initial state where two scalar fields are entangled by following the research~\cite{Albrecht:2014aga}, in which it was argued that oscillatory spectra would appear due to the entanglement. They considered an initially four mode squeezed state of the two free massive scalar fields $\phi_{\bm k}$, $\chi_{\bm k}$, one of them is assumed to be an inflaton field, in de Sitter space in the Schr\"{o}dinger formalism. Since in their Schr\"odinger formalism it is not easy to see the physical origin of the oscillatory behavior in the spectrum, we instead adopted the Heisenberg formalism. In the Heisenberg formalism, the state considered in~\cite{Albrecht:2014aga} is described
by a Bogoliubov transformation that mixes a positive frequency function of $\phi_{\bm k}$ with a negative frequency function of $\chi_{\bm k}$. We then calculated the exact power spectrum for the $\phi$ field and clarified that the origin of an oscillatory spectrum has to be in the absolute value of the Bogoliubov coefficients $|\beta_k|$.

To confirm that our state in the Heisenberg picture is the same as the state studied in~\cite{Albrecht:2014aga}, we derived the wavefunction for our state in the Schr\"odinger picture in the same form as assumed in~\cite{Albrecht:2014aga}. We note that the resultant wavefunction we obtained is exact and general for the class of entanglement considered in~\cite{Albrecht:2014aga}. To show explicitly the identity between an expression for the
spectrum given in terms of functions appearing in the wavefunction as given in Eq.~(\ref{exact2}) and the exact expression we obtained, we evaluated the former approximately by expanding it in terms of the Bogoliubov coefficients $\beta_k$. We found that the spectrum expanded up to the second order of the Bogoliubov coefficients $\beta_k$ completely agrees with the exact spectrum we derived in the Heisenberg formalism. From those analyses, we conclude that oscillatory spectra due to the initial state entanglement can be produced only when the absolute value of the Bogoliubov coefficient $|\beta_k|$ is chosen to have an oscillatory feature in $k$.

We here stress that, on the contrary, an initially non-entangled state may naturally produce an oscillatory spectrum due to quantum interference if the initial state deviates
from the Bunch-Davies vacuum as shown in~\cite{Kanno:2015lja}. For example, if we consider a squeezed
state of $\phi_{\bm k}$ given by a Bogoliubov transformation of the Bunch-Davies
vacuum, an oscillatory spectrum appears if the relative phase between
the coefficients $\alpha_k$ and $\beta_k$ depends on $k$ even if there
is no oscillation in the absolute value of $\beta_k$. Thus we should not be confused the oscillatory spectra as an effect of entanglement. In order to see solely the effect of entanglement in the spectrum, we should rather see the effect of quantum interference that may distinguish the spectrum of initially entangled state from the one of non-entangled state~\cite{Kanno:2015lja}.

\section*{Acknowledgments}
I would like to thank Misao Sasaki and Jiro Soda for valuable discussions, useful suggestions and comments. I am grateful for the hospitality by Kobe University where this work was completed. This work was supported by IKERBASQUE, the Basque Foundation for
Science.

\end{document}